\documentclass[5pt]{article}

\usepackage[letterpaper]{geometry}
\usepackage{spconf,amsmath,epsfig}
\usepackage{enumitem}

\usepackage{fancyhdr}

\usepackage{multirow}
\usepackage{booktabs}
\usepackage{threeparttable}
\usepackage{xcolor}

\usepackage{graphicx}
\usepackage{subfigure}

\begin{document}\sloppy

\thispagestyle{fancy}
\fancyhf{} 
\renewcommand{\headrulewidth}{0pt}

\pagestyle{empty}

\def\x{{\mathbf x}}
\def\L{{\cal L}}

\title{Quality Assessment for Tone-Mapped HDR Images Using Multi-scale and Multi-layer Information}
%

\name{Qin He$^1$, Dingquan Li$^{1,3}$, Tingting Jiang$^{1,2}$, Ming Jiang$^{1,3}$}
\address{
$^1$National Engineering Lab for Video Technology, Cooperative Medianet Innovation Center; \\  
$^2$School of EECS; 
$^3$LMAM, SMS \& BICMR, 
Peking University, Beijing 100871, China \\
heqin6314@gmail.com, \{dingquanli, ttjiang, ming-jiang\}@pku.edu.cn
}

\maketitle

\begin{abstract}
Tone mapping operators and multi-exposure fusion methods allow us to enjoy the informative contents of high dynamic range (HDR) images with standard dynamic range devices, but also introduce distortions into HDR contents.
Therefore methods are needed to evaluate tone-mapped image quality. Due to the complexity of possible distortions in a tone-mapped image, information from different scales and different levels should be considered when predicting tone-mapped image quality. 
So we propose a new no-reference method of tone-mapped image quality assessment based on multi-scale and multi-layer features that are extracted from a pre-trained deep convolutional neural network model. After being aggregated, the extracted features are mapped to quality predictions by regression. The proposed method is tested on the largest public database for TMIQA and compared to existing no-reference methods. The experimental results show that the proposed method achieves better performance.
\end{abstract}
\section{Introduction}
\label{sec:intro}

High dynamic range (HDR) images, which provide wider range of luminance variation draw more attention in recent years.
But it is still not accessible to most people who use traditional standard dynamic range (SDR) devices.
Tone mapping operators and multi-exposure fusion methods~\cite{HDRbook} provide alternative approaches to access HDR contents with SDR devices.
The resulting images, called tone-mapped HDR images, can preserve the details and the contrast of HDR contents in a narrower dynamic range. But these methods also bring complex distortions into the HDR contents in the practical applications~\cite{Attributes}.
Therefore, proper image quality assessment (IQA) methods are necessary.
Ideally, the best solution is subjective assessment carried by human beings, but subjective assessment is complicated and time-consuming.
Thus proper objective assessment methods are needed to predict tone-mapped image quality automatically.
Furthermore, the absence of reference HDR images calls for no-reference image quality assessment (NR-IQA) methods, since tone-mapped images are designed for situations where HDR devices are not accessible.

\begin{figure}
\centering 
\subfigure[MOS=18.3126018946] { \label{fig:a} 
\includegraphics[width=0.45\columnwidth]{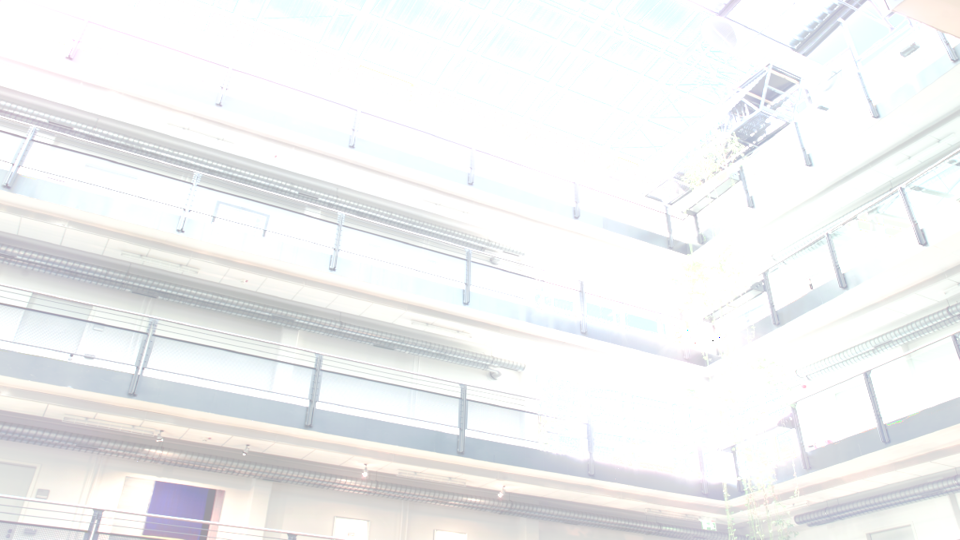} 
} 
\subfigure[MOS=51.3429624349] { \label{fig:b} 
\includegraphics[width=0.45\columnwidth]{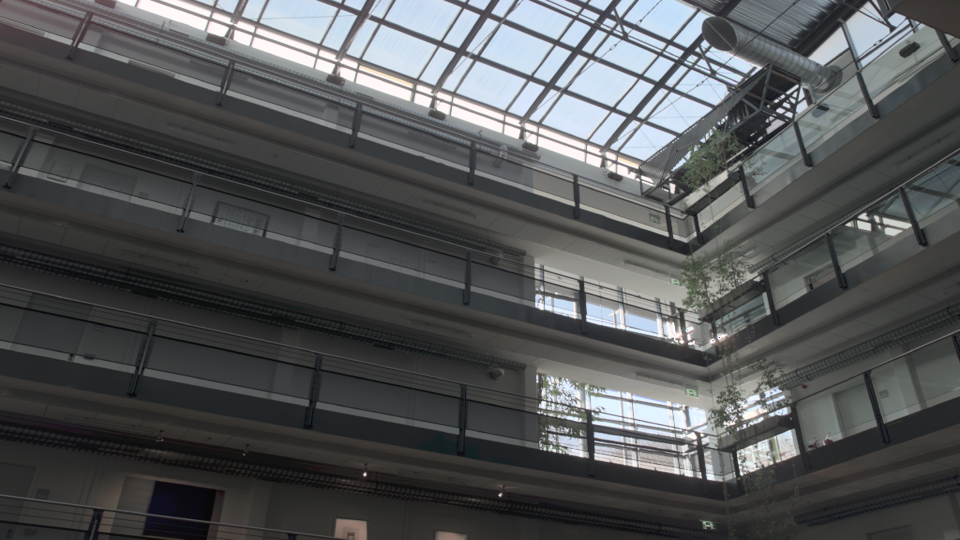}
} 
\subfigure[MOS=54.4365059326] { \label{fig:c} 
\includegraphics[width=0.45\columnwidth]{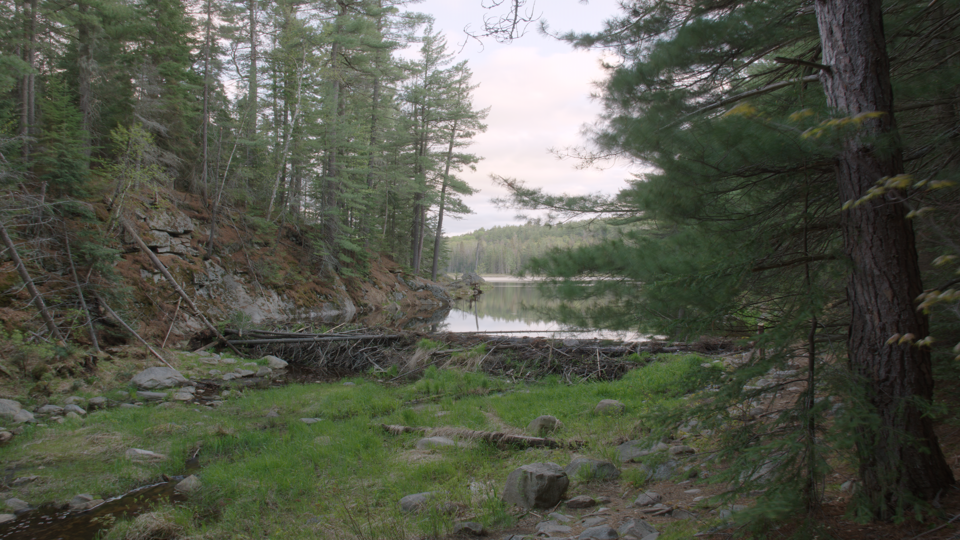} 
} 
\subfigure[MOS=23.6067858845] { \label{fig:d} 
\includegraphics[width=0.45\columnwidth]{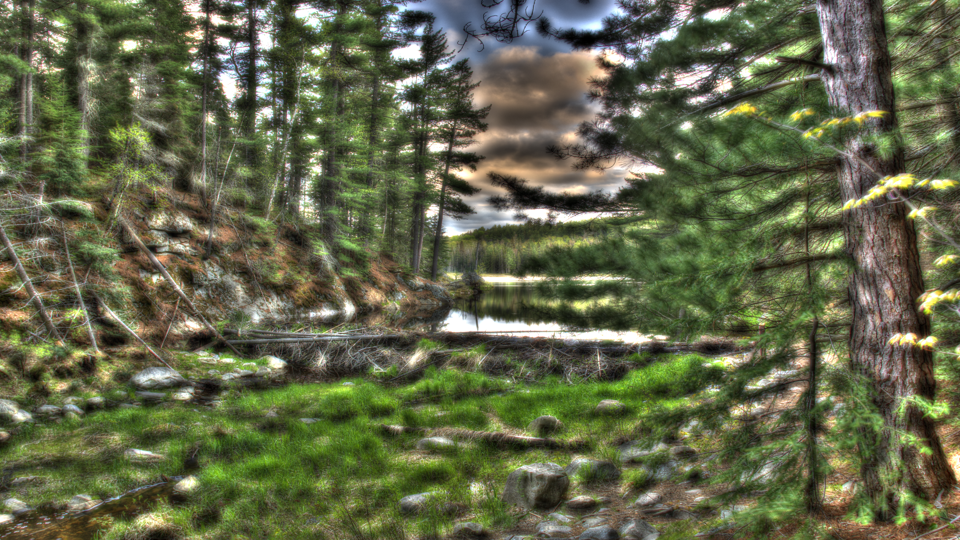} 
} 
\caption{Four images from ESPL-LIVE HDR Database. Higher MOS indicates better subjective image quality.} 
\label{fig:0} 
\end{figure}





However, the aim to preserve HDR contents on SDR devices also leads to the complex standard in the quality assessment of tone-mapped HDR images.
Normally, basic attributes like brightness and contrast are important to IQA because they can be used to measure the preservation of details and structures.
Images with moderate brightness and higher local contrast usually have more information and better image quality.
It is shown in the comparison between Fig.~\ref{fig:0} (a) and Fig.~\ref{fig:0} (b).
Focusing on the details and contrast preservation, existing objective assessment methods mainly take structural similarity and natural scene statistics as tools of tone-mapped image quality assessment (TMIQA).
But due to the complexity of TMIQA,
more factors should be considered.
Decided by the aim to preserve HDR contents, tone-mapped images should have more information and higher contrast than general images.
But then they can never preserve all the information from the original HDR images because of the reduction of dynamic range.
The proper level of preservation depends on the overall scene and the contents in the image.
For example, we can see that compared with Fig.~\ref{fig:0} (c), Fig.~\ref{fig:0} (d) preserves more details and color information in the sky and in the shadow of the woods.
But preserving higher local contrast also breaks the balance
between the sky and the woods and affects the recognition of the scene.
Without high level information and global information,
the image quality cannot be predicted using only low-level information and local information.

In this paper we propose a new no-reference image quality assessment (NR-IQA) method to evaluate tone-mapped image quality which makes use of multi-scale and multi-layer information, adapted to the complex standard of TMIQA.
A multi-scale representation is adopted to combine local and global information.
Multi-Layer features are extracted from different layers of a pre-trained deep convolutional neural network (DCNN) model, to give consideration to both low and high level information.
Then these informative features are mapped to quality prediction by regression.
We conduct experiments on the largest public available tone-mapped image database. The experimental results prove that our proposed method outperforms existing NR-IQA methods on quality assessment for tone-mapped images.

The paper is organized as follow. Section 2 reviews related research. Section 3 introduces the ESPL-LIVE HDR database. The proposed method is described in Section 4. Experimental results are given in Section 5, and lastly conclusions are made in Section 6.

\section{Related Work}

The preservation of information in extra dynamic range, the balance between local and global contrast, and the natural looking are important aspects of tone-mapped image quality.
Therefore, existing methods mainly access tone-mapped image quality from detail preservation, contrast preservation, and image naturalness.
Compared to the employment of HDR and tone mapping operators, research history of TMIQA is relatively short.
As far as we concerned, only a few objective assessment methods are proposed.
According to the availability and usage of original HDR images, they can be further classified as full-reference (FR) methods and no-reference (NR) methods.

\subsection{Full-Reference Methods}
 Due to the importance of detail preservation, many researchers naturally take structural similarity between the original HDR images and the tone-mapped ones as a measure of tone-mapped image quality.
 Aydin et al.~\cite{08SIGGRAPH} first break the limitation of different dynamic range with proper transformation and measure the similarity between HDR and reproduced SDR images.
 Yeganeh et al.~\cite{13TIP} propose the popular FR framework of Tone-Mapped image Quality Index (TMQI), which contains two components: multi-scale local structural similarity to evaluate the information preservation, and statistical naturalness to evaluate the natural looking of tone-mapped images.
 Many modified methods are proposed on the basis of TMQI.
 Nasrinpour et al.~\cite{15ICIP} use visual saliency, while Xie et al.~\cite{16ISCAS} choose sparse domain instead, to modify the measurement of local structural similarity;
 Kundu et al.~\cite{16ICIP} use the statistical features of mean subtracted contrast normalized (MSCN) pixels to measure the natural statistics, and choose a pooling strategy from standard deviation of MSCN pixels, visual saliency and local information entropy for the pooling of local structural similarity.

\subsection{No-Reference Methods}
 Researches on the contributions of various image attributes to tone-mapped image quality are carried by {\v{C}}ad{\'\i}k et al.~\cite{Attributes} and show interesting results that inherent attributes such as brightness, contrast and color contribute even more to tone-mapped image quality than the information preservation.
 Thus TMIQA can be done using these attributes without information from the reference images. 
 
 Since structural similarity is not applicable, no-reference methods of TMIQA focus on the design and extraction of various features to deal with the complex distortions in tone-mapped images, and use machine learning methods to assess the quality.
 For example, Kundu et al.~\cite{17TIP}
 extract 40 statistical features such as mean, standard deviation, skewness and kurtosis from spatial domain, gradient domain, standard deviation field of MSCN pixels and 6 types of log-derivative domains, and implement $\epsilon$-SVR  on these features to predict the quality of tone-mapped images.
 Yue et al.~\cite{17TIE} adopt a human visual system inspired method, where tone-mapped images are decomposed into four pairs of opponent color channels, and the features of local structures in the simulated responses of Single-Opponent/Double-Opponent cells are used to train models for TMIQA.
 
 However, the structural information in the tone-mapped images is still accessible without reference HDR images.
 Inspired by the goal to preserve details in the extra dynamic range,
 Gu et al.~\cite{16TMM} replace structural similarity component in the TMQI framework with the mean of Sobel map of each image,
 while the information entropy of the brightened and darkened versions of the image is used to quantify the volume of details.
 
Existing FR-TMIQA methods based on TMQI make use of multi-scale structural features but neglect the high-level information in the tone-mapped images.
Existing NR-TMIQA methods focus on the design of different features, but the diversity of information is limited by the number of features.
Unlike existing methods, our method use the pre-trained DCNN model to extract different levels of features for quality assessment.
Furthermore both local and global features are considered by using multi-scale representations.
These steps adapt our method to the characteristic of TMIQA.

\begin{figure*}[htb]
\centering\includegraphics[width=5.8in]{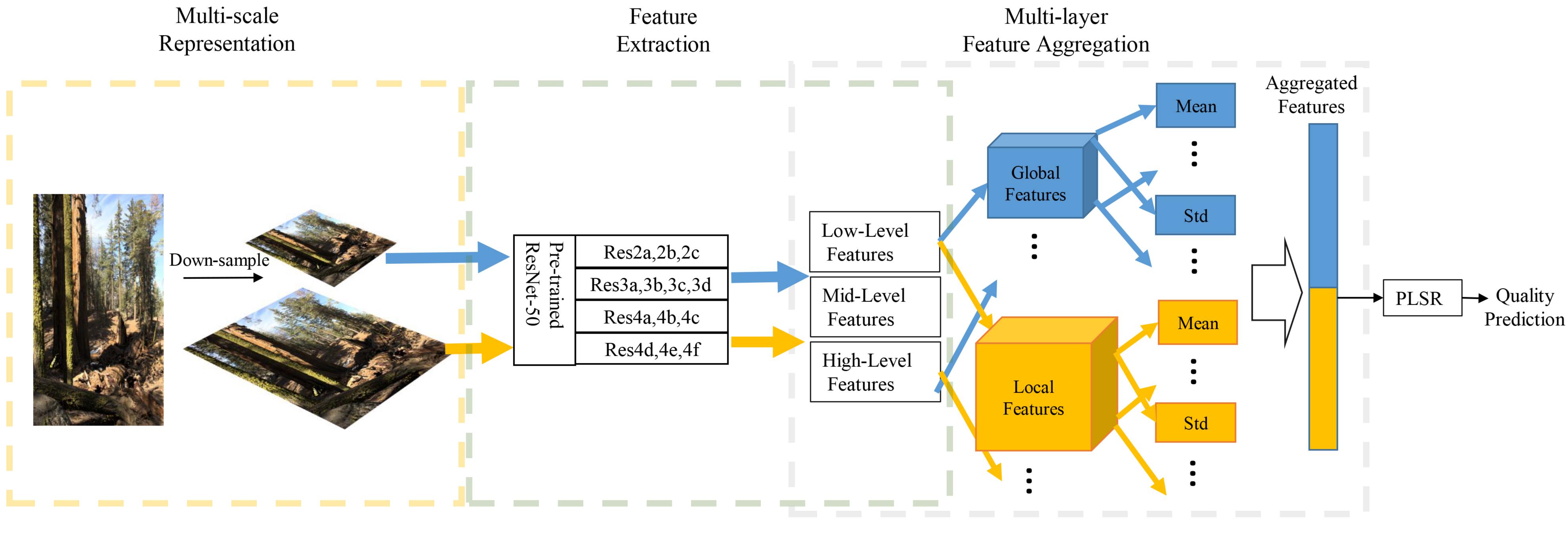}
\caption{The framework of proposed method. Features with necessary information are guaranteed mainly by multi-scale representation and multi-layer feature aggregation.}\label{fig:1}
\end{figure*}

\section{Database}

 To investigate the best proposal of our method, and to test its performance, we take the newest public available tone-mapped HDR image database, ESPL-LIVE HDR Database~\cite{ESPL} for experiments.
 So far it is the largest database for TMIQA,
 and it covers a greater diversity of methods to obtain tone-mapped images:
 images in the database are derived from HDR illuminance maps and SDR image stacks through 11 different methods.
 Therefore it is suitable for experiments of our method, which is aimed at predicting tone-mapped HDR image quality affected by complex factors.

 There are 1811 tone-mapped HDR images in total,
 and the images fall into three types by the ways they are obtained:
 747 images are created
 through 4 different tone mapping operators;
 710 images are created directly from SDR image stacks through 5 multi-exposure fusion methods;
 354 images are obtained by Photomatix with 2 different post-processing settings.
 The images have been evaluated by 5462 participants on a crowdsourcing platform, and the subjective sores are provided in the form of mean opinion scores (MOS) ranging from 0 to 100.
 
 We randomly split the data into training and testing sets at 4:1 ratio in the objective experiments which will be introduced in  Section~\ref{sec:exp}.
 20\% of the training sets are further taken as validation sets in the validation experiments which will be introduced in Section~\ref{sec:method}, to investigate proposals of our method.
 Care was taken to ensure the data independency between training, validation and testing sets.


\section{Proposed Method}
\label{sec:method}
As stated in Section~\ref{sec:intro}, both low-level and high-level information, as well as global and local information should be considered in TMIQA.
To describe information of this large span with designed features is difficult.
Therefore we decide to extract features
with the aid of the DCNN model.
Databases for IQA are relatively small owing to the expense of subjective assessment.
It is difficult to train a model deep enough to extract high-level features from scratch.
Inspired by the work of Li et al.~\cite{exploiting}, we choose the ResNet-50 model~\cite{resnet} trained on ImageNet as the pre-trained model for feature extraction.
On one hand, due to the variety in image quality and the large number of images in ImageNet, the model which is pre-trained on ImageNet is a good choice to extract features for IQA.
On the other hand, ResNet-50 is deep enough to extract high-level features we need, while preserving some primary information related to image quality in residual images.

To give proper quality assessment to tone-mapped HDR images, the proposed method consists of four parts: multi-scale image representation, feature extraction, multi-layer feature aggregation, and quality prediction.
\textcolor{black}
{Firstly an image is represented with its original and down-sampled version and fed into the modified ResNet-50,
to preserve both local and global information.
Then feature maps with information of different levels are extracted from selected layers of the pre-trained model.
After being extracted, feature maps are aggregated through mean and standard deviation pooling in each channel, and concatenated according to the information carried, to form a feature structure to clearly preserve the information for the final mapping.}
Finally, the multi-scale and multi-layer features are mapped to one single quality score by partial least squares regression (PLSR)~\cite{PLSR}.

The whole framework is shown in Fig.~\ref{fig:1}. Details of multi-scale representation and multi-layer feature aggregation will be discussed below.

\subsection{Multi-scale image representation and feature extraction}

One important role of the tone-mapped image is to preserve the contrast and the details in dark and highlight areas.
Therefore both local information that can be used to evaluate the detail preservation, and global information that can be used to evaluate the contrast preservation, are important to TMIQA.
Features with local information can be easily extracted from the pre-trained DCNN model, since the input size of ResNet-50 is smaller than the size of images in ESPL-LIVE HDR Database.
But most of the global information may be lost, if the image is cropped to fit the model.
Therefore we remove the fully-connected layer and consequently remove the limitation of fixed input size.
And feature maps will be extracted when the size of the input image is larger than the original input size of the pre-trained model.
However, these feature maps still contains little global information without fully-connected layer.
Because the receptive field of one neuron in ResNet-50 is at most  $32 \times 32$ after modified and that is much smaller than the image size.
As a result, one element in the output feature maps is mainly decided by a small area.
To preserve global information within this fixed receptive field, we take the down-sampled version of an image as one component of the image representation, to allow larger areas and more information being involved in the first place.
The down-sampling operation is done along both height and width with a factor of 2.
Furthermore, to avoid the impact of down-sampling operation on image quality and to make use of local information, the original image is also preserved and fed into the DCNN model.

Feature maps extracted from images of different scales will be in different size.
For example, an image with size of $960 \times 540 \times 3$ is fed into the modified ResNet-50, the feature maps extracted from layer $res4f$ will be $30 \times 16 \times 1024$, while its down-sampled version gets feature maps sized $15 \times 8 \times 1024$.
We implement mean and standard deviation pooling on each channel of the feature maps, and the results from all the channels are taken as the representative statistical features of each scale.
After that, features of the two scales share the same size of $2 \times 1 \times 1024$.
They are then simply concatenated to form multi-scale features to preserve the full and clear structure of the multi-scale representation.
After down-sampling, the multi-scale features of an image can be described as $$O \oplus D,$$  where $O$ is features from the original image, $D$ is the features from the image after down-sampling operation, $\oplus$ is the concatenation operator.

To validate the gain of multi-scale representation, we compare the performance of the multi-scale representation with that of the single-scale representation where only the original image is involved.
We take features from layer $res2c, res3d$ and $res4f$, as the representative features of different levels,
and then test the performance of different representations.
The average performance of the two representations is also reported.
The comparison is done on the validation sets from ESPL-LIVE HDR Database~\cite{ESPL}, and the results are reported in Table~\ref{tab:multi-scale}.
It can be seen that the multi-scale representation achieves better performance.
Therefore, in our method, we choose the multi-scale representation to preserve both local and global information of the tone-mapped HDR images.

\begin{table}[!htb]
\begin{center}
\caption{Comparison between the performance of the multi-scale representation and the single-scale representation. Medians of SROCC in 1000 runs are reported.} \label{tab:multi-scale}
\vspace{2mm}
\resizebox{0.9\columnwidth}{!}
{
\begin{tabular}{ccccc} 
\toprule  
Features & res2c & res3d & res4f & overall \\  
\midrule  $O$ & 0.709 & 0.763 & 0.787 & 0.753 \\  
$O \oplus D$ & \textbf{0.712} & \textbf{0.767} & \textbf{0.796} & \textbf{0.758} \\  
\bottomrule
\end{tabular}
}
\end{center}
\end{table}

\subsection{Multi-layer feature aggregation}
\label{subsec:mulilayer}
 As stated in Section~\ref{sec:intro}, tone-mapped images should preserve more details and contrast while not affecting the contents in the original images.
 When the reference images are not available, it is difficult to estimate the proper level of detail and contrast preservation without high-level information, while basic attributes such as brightness and contrast are important as well.
 Therefore it can be expected that tone-mapped image quality can be better predicted with features extracted from different layers, i.e. multi-layer features.
 
 To construct these multi-layer features, we combine features from different layers of ResNet-50.
 Based on the core algorithm of residual learning, the main part of ResNet-50 can be divided into 16 building blocks.
 Each building block consists of multiple layers and only the ending layer outputs the result of residual learning.
 Thus we give consideration to these 16 ending layers of the blocks as the layers to extract features from.
 Layers after $res4f$ are ignored, which are aimed at image classification but may have less relation to TMIQA.
 The rest 13 layers are further divided into four sets according to the depth: \{$res2a, res2b, res2c$\}, \{$res3a, res3b, res3c, res3d$\}, \{$res4a, res4b, res4c$\}, \{$res4d, res4e, res4f$\}.
 Three layers are chosen to extract low-level, mid-level and high-level features, where at most one layer is chosen for one set in order to achieve a balance of low dimensions and diversity of features. 
 Features from the three selected layers are then concatenated as multi-layer features in the form of $$f_l \oplus f_m \oplus f_h,$$ where $f_l$, $f_m$, $f_h$ are the low-level, the mid-level and the high-level features separately.
 And $\oplus$ is the concatenation operator, adopted to preserve the clear structure of the multi-layer representation.

\begin{table}[!tb]
\begin{center}
\caption{Comparison between the performance of the single layer features and the multi-layer features, where TM stands for Tone Mapping, MEF stands for Multi-Exposure Fusion, and PP stands for Post Processing. Medians of SROCC in 1000 runs are reported.}
\label{tab:multi-layer}
{\begin{tabular}{ccccc}  
\toprule  
Features & Overall & TM & MEF & PP   \\  
\midrule  
$f_l$ (res2a) & 0.712 & 0.786 & 0.654 & 0.576\\  
$f_m$ (res4b) & 0.800 & 0.845 & 0.755 & 0.729\\  
$f_h$ (res4f) & 0.796 & 0.838 & 0.754 & 0.722\\  
$f_l \oplus f_m \oplus f_h$ & \textbf{0.814} & \textbf{0.854} & \textbf{0.771} & \textbf{0.744}\\  
\bottomrule
\end{tabular}}
\end{center}
\end{table}

 Experiments are done to choose the combination of layers with the best performance.
 We choose three sets from all the four sets, and each set has three or four layers to choose.
 So there are 135 combinations of layers in total.
 We compare the performance of all the combinations by running experiments 1000 times on the validation sets from ESPL-LIVE HDR Database, and find that the combination of layer $res2a, res4b, res4f$ achieves the best validation performance among all the 135 combinations.
 Furthermore, the same experiments are also done with the single layer features to validate the gain of the multi-layer features.
 Results of the comparison between the features from the selected combination and from the single layers in the combination are reported in Table~\ref{tab:multi-layer}.
 As what we expect, multi-layer features performs better than the single layer features.
 

 Features are adapted for TMIQA after aggregation.
 However, the length of the multi-scale and multi-layer features after aggregation is $$l_A = (l_{f_l} + l_{f_m} + l_{f_h}) \times 2,$$ where $l_A, l_{f_l}, l_{f_m}, l_{f_h}$ are the length of the aggregated features and features from the single layers.
 The dimension of features we adopted is 4608, which is much larger than the number of training samples.
 Features of such dimension may cause overfitting in regression.
 Therefore, we adopt partial least squares regression (PLSR)~\cite{PLSR} in our work, because it is suitable for regression tasks which have more variables than observations.
 PLSR reduces the dimension of input features to a few unrelated latent components, then maps these components to quality predictions.
 The number of latent components in PLSR is selected from the range $[10,20]$ by running experiments 1000 times on the validation sets,
 and the results show that the PLSR model with 15 latent components achieves the best performance.
 
 After investigations in this section, we obtain the best proposal for our method.
 Tone-mapped images are down-sampled for once, and the original and the down-sampled version of the images are fed into the pre-trained DCNN model of modified ResNet-50.
 Then feature maps are extracted from layer $res2a, res4b$ and $res4f$, then pooled by the mean and standard deviation of each channel, and concatenated to form the multi-scale and multi-layer features.
 The aggregated features are mapped to quality scores by the PLSR model with 15 latent components.

 \begin{table}[!tb]
\centering	
\caption{Medians of SROCC, PLCC and RMSE between quality predictions and the MOS scores on the ESPL-LIVE HDR database. The best two values in each column are marked in boldface.}  
\vspace{2mm}
\label{tab:result2}	
\begin{small}	
\begin{threeparttable}	
\begin{tabular}{cp{16mm}p{13mm}p{13mm}}  	
\toprule	IQA & SROCC & PLCC & RMSE\\  
\midrule
Proposed(100 runs)&\textbf{0.823}&\textbf{0.827}&\textbf{5.697}\\ 
Proposed(1000 runs)&\textbf{0.820}&\textbf{0.824}&\textbf{5.768}\\ 
HIGRADE-2~\cite{17TIP} & 0.730 & 0.728 & 6.992\\
BIBQA~\cite{17TIE}     & 0.702 & 0.692 & 7.133\\
DESIQUE~\cite{DESIQUE} & 0.570 & 0.568 & 8.296\\    
GM-LOG~\cite{GMLOG}    & 0.556 & 0.557 & 8.357\\    
BRISQUE~\cite{MSCN}    & 0.418 & 0.444 & 9.049\\
\bottomrule	
\end{tabular}
\begin{tablenotes}   
\footnotesize 
\item[] Results of HIGRADE-2, DESIQUE, GM-LOG and BRISQUE are quoted from~\cite{17TIP}.
Results of BIBQA are quoted from~\cite{17TIE}.   
\end{tablenotes}	
\end{threeparttable}	
\end{small}
\end{table}

\section{Experiments}
\label{sec:exp} 
 
In this section, we conduct experiments on ESPL-LIVE HDR Database~\cite{ESPL} to compare the performance of our proposed method with some existing TMIQA methods and some state-of-the-art NR-IQA methods which are aimed at general SDR images.
Because most of the existing TMIQA methods are full-referenced and the original HDR images are not available in the ESPL-LIVE HDR Database, we compared two NR-TMIQA methods among them.
The compared NR-TMIQA methods are HIGRADE method proposed by Kundu et al.~\cite{17TIP}, and 
the biologically inspired blind quality assessment (BIBQA) method proposed by Yue et al.~\cite{17TIE}.
Other general NR-IQA methods compared are DESIQUE~\cite{DESIQUE}, GM-LOG~\cite{GMLOG}, and BRISQUE~\cite{MSCN}.
The proposed method is implemented on MATLAB platform.
The results of HIGRADE, DESIQUE, GM-LOG, BRISQUE on the same database are quoted from~\cite{17TIP}, and the result of BIBQA method are quoted from~\cite{17TIE}.
However, the experimental settings are different in these two cited papers. In~\cite{17TIP}, the data is randomly split into training and testing sets at 4:1 ratio for 100 times; in~\cite{17TIE}, the data is split for 1000 times at the same ratio. 
To make the comparison fair, we conduct experiments on both settings for our method.
\textcolor{black}
{
Three commonly used criteria are adopted to evaluate the performance:
Spearman's rank correlation coefficient (SROCC), which can demonstrate the prediction monotonicity,
Pearson's linear correlation coefficient (PLCC) and
root mean-squared error (RMSE), which are used for measuring prediction accuracy.
These criteria are calculated for every run,
and the medians of the three criteria in 100 runs and 1000 runs are reported separately. 
}
Table~\ref{tab:result2} reports the comparison results.
Performance of the best proposal of HIGRADE method, HIGRADE-2 is reported as the representative performance of HIGRADE method.
And the medians in 1000 runs are quoted from \cite{17TIE} as the result of BIBQA, because the data is randomly split and there shouldn't be obvious difference between the medians in 100 or 1000 runs.
As reported in the table, comparison results prove that our proposed method achieves better performance.

\section{Conclusion}

In this paper, we propose a new NR-IQA method for tone-mapped HDR images. Multi-scale and multi-layer features are extracted from the pre-trained ResNet-50, and mapped to quality predictions by partial least squares regression. The performance of our proposed method is tested on the largest database for tone-mapped HDR image assessment, ESPL-LIVE HDR Database, and compared to the existing NR-IQA methods aimed at tone-mapped images and general images. The experimental results show that our method outperforms all the existing NR-IQA methods on tone-mapped image quality assessment, and also prove that multi-scale and multi-layer features lead to performance improvements in tone-mapped image quality assessment.

\section{Acknowledgment}
    This work was partially supported by National Basic Research Program of China (973 Program) under contract 2015CB351803 and the Natural Science Foundation of China under contracts 61572042, 61390514, 61527804.
    We also acknowledge the High-Performance Computing Platform of Peking University for providing computational resources.

\bibliographystyle{IEEEbib}
\bibliography{camera-ready_icme2018template}

\begin{thebibliography}{10}

\bibitem{HDRbook}
Erik Reinhard, Wolfgang Heidrich, Paul Debevec, Sumanta Pattanaik, Greg Ward,
  and Karol Myszkowski,
\newblock {\em High Dynamic Range Imaging: Acquisition, Display, and
  Image-Based Lighting},
\newblock Morgan Kaufmann, 2010.

\bibitem{Attributes}
Martin {\v{C}}ad{\'\i}k, Michael Wimmer, Laszlo Neumann, and Alessandro Artusi,
\newblock ``Image attributes and quality for evaluation of tone mapping
  operators,''
\newblock in {\em Pacific Conference on Computer Graphics and Applications},
  2006, pp. 35--44.

\bibitem{08SIGGRAPH}
Tun{\c{c}}~Ozan Aydin, Rafa{\l} Mantiuk, Karol Myszkowski, and Hans-Peter
  Seidel,
\newblock ``Dynamic range independent image quality assessment,''
\newblock {\em ACM Trans. Graphics}, vol. 27, no. 3, pp. 69, 2008.

\bibitem{13TIP}
Hojatollah Yeganeh and Zhou Wang,
\newblock ``Objective quality assessment of tone-mapped images,''
\newblock {\em IEEE Trans. Image Process.}, vol. 22, no. 2, pp. 657--667, 2013.

\bibitem{15ICIP}
Hamid~Reza Nasrinpour and Neil~DB Bruce,
\newblock ``Saliency weighted quality assessment of tone-mapped images,''
\newblock in {\em IEEE ICIP}, 2015, pp. 4947--4951.

\bibitem{16ISCAS}
Lijuan Xie, Xiang Zhang, Shiqi Wang, Xinfeng Zhang, and Siwei Ma,
\newblock ``Quality assessment of tone-mapped images based on sparse
  representation,''
\newblock in {\em IEEE ISCAS}, 2016, pp. 2218--2221.

\bibitem{16ICIP}
Debarati Kundu and Brian~L Evans,
\newblock ``Visual attention guided quality assessment of tone-mapped images
  using scene statistics,''
\newblock in {\em IEEE ICIP}, 2016, pp. 96--100.

\bibitem{17TIP}
Debarati Kundu, Deepti Ghadiyaram, Alan~C Bovik, and Brian~L Evans,
\newblock ``No-reference quality assessment of tone-mapped {HDR} pictures,''
\newblock {\em IEEE Trans. Image Process.}, vol. 26, no. 6, pp. 2957--2971,
  2017.

\bibitem{17TIE}
Guanghui Yue, Chunping Hou, Ke~Gu, Shasha Mao, and Wenjun Zhang,
\newblock ``Biologically inspired blind quality assessment of tone-mapped
  images,''
\newblock {\em IEEE Trans. Ind. Electron.}, 2017.

\bibitem{16TMM}
Ke~Gu, Shiqi Wang, Guangtao Zhai, Siwei Ma, Xiaokang Yang, Weisi Lin, Wenjun
  Zhang, and Wen Gao,
\newblock ``Blind quality assessment of tone-mapped images via analysis of
  information, naturalness, and structure,''
\newblock {\em IEEE Trans. Multimedia}, vol. 18, no. 3, pp. 432--443, 2016.

\bibitem{ESPL}
Debarati Kundu, Deepti Ghadiyaram, Alan Bovik, and Brian Evans,
\newblock ``Large-scale crowdsourced study for tone mapped {HDR} pictures,''
\newblock {\em IEEE Trans. Image Process.}, vol. 26, no. 10, pp. 4725--4740,
  2017.

\bibitem{exploiting}
Dingquan Li, Tingting Jiang, and Ming Jiang,
\newblock ``Exploiting high-level semantics for no-reference image quality
  assessment of realistic blur images,''
\newblock in {\em ACM on Multimedia Conference}, 2017, pp. 378--386.

\bibitem{resnet}
Kaiming He, Xiangyu Zhang, Shaoqing Ren, and Jian Sun,
\newblock ``Deep residual learning for image recognition,''
\newblock in {\em IEEE CVPR}, 2016, pp. 770--778.

\bibitem{PLSR}
Roman Rosipal and Nicole Kr{\"a}mer,
\newblock ``Overview and recent advances in partial least squares,''
\newblock in {\em Subspace, Latent Structure and Feature Selection}, pp.
  34--51. 2006.

\bibitem{DESIQUE}
Yi~Zhang and Damon~M. Chandler,
\newblock ``No-reference image quality assessment based on log-derivative
  statistics of natural scenes,''
\newblock {\em Journal of Electronic Imaging}, vol. 22, pp. 22 -- 22 -- 23,
  2013.

\bibitem{GMLOG}
Wufeng Xue, Xuanqin Mou, Lei Zhang, Alan~C Bovik, and Xiangchu Feng,
\newblock ``Blind image quality assessment using joint statistics of gradient
  magnitude and {Laplacian} features,''
\newblock {\em IEEE Trans. Image Process.}, vol. 23, no. 11, pp. 4850--4862,
  2014.

\bibitem{MSCN}
Anish Mittal, Anush~Krishna Moorthy, and Alan~Conrad Bovik,
\newblock ``No-reference image quality assessment in the spatial domain,''
\newblock {\em IEEE Trans. Image Process.}, vol. 21, no. 12, pp. 4695--4708,
  2012.

\end{thebibliography}

\end{document}